\documentclass[useAMS,usenatbib]{mn2e}

\usepackage{graphicx}

\def\jh{\mbox{$\rm (J-H)$}}

\def\jk{\mbox{$\rm (J-K_s)$}}

\def\ebv{\mbox{$\rm E(B-V)$}}
\def\ejh{\mbox{$\rm E(J-H)$}}
\def\rc{\mbox{$\rm R_{C}$}}
\def\rl{\mbox{$\rm R_{lim}$}}
\def\rt{\mbox{$\rm R_{t}$}}
\def\ms{\mbox{$\rm M_\odot$}}
\def\ds{\mbox{$\rm d_\odot$}}
\def\tdis{\mbox{$\rm t_{dis}$}}

\def\Rgc{\mbox{$\rm R_\odot$}}
\def\dgc{\mbox{$\rm R_{GC}$}}
\def\rx{\mbox{$\rm R_{ext}$}}

\def\jj{\mbox{$\rm J$}}
\def\hh{\mbox{$\rm H$}}
\def\ks{\mbox{$\rm K_s$}}
\def\aV{\mbox{$\rm A_V$}}
\def\ns{\mbox{$\rm N_{1\sigma}$}}
\def\no{\mbox{$\rm N_{obs}$}}
\def\nc{\mbox{$\rm N_{cl}$}}
\def\sFS{\mbox{$\rm\sigma_{FS}$}}

\title[Star cluster candidates in central directions]{Probing FSR star cluster candidates
in bulge/disk directions with 2MASS colour-magnitude diagrams}

\author[Bica et al.]{E. Bica$^1$, C. Bonatto$^1$ and D. Camargo$^1$\\
$^1$ Departamento de Astronomia, Universidade Federal do Rio Grande do Sul\\
Av. Bento Gon\c{c}alves 9500, Porto Alegre 91501-970, RS, Brazil; charles@if.ufrgs.br;
bica@if.ufrgs.br}

\begin{document}


\maketitle


\begin{abstract}
We analyse 20 star cluster candidates projected mostly in the bulge direction ($|\ell|<60^\circ$).
The sample contains all candidates in that sector classified by \citet{FSRcat} with quality flags
denoting high probability of being star clusters. Bulge contamination in the colour-magnitude
diagrams (CMDs) is in general important, while at lower Galactic latitudes disk stars contribute
as well. Properties of the candidates are investigated with 2MASS CMDs and stellar radial density
profiles (RDPs) built with field star decontaminated photometry. To uncover the nature of the
structures we decontaminate the CMDs from field stars using tools that we previously developed to
deal with objects in dense fields. We confirm in all cases excesses in the RDPs with respect to 
the background level, as expected from the method the candidates were originally selected. CMDs 
and RDPs taken together revealed 6 open clusters, 5 uncertain cases that require deeper observations, 
while 9 objects are possibly field density fluctuations.
\end{abstract}


\section{Introduction}
\label{intro}

On a broad perspective, any self-gravitating group of stars whose members share common initial conditions
can be classified as a star cluster. In this definition fit the embedded, open and globular clusters (OCs
and GCs, respectively), as well as the OC remnants. Those objects span a wide range of ages, masses and
luminosities, among other parameters. While the upper limit to the GC population may be close to 200 members
(e.g. \citealt{FSR1767}), the OC (as well as embedded and remnants) census, which at this moment amounts to
more than $\sim1000$ according to the WEBDA\footnote{\em obswww.univie.ac.at/webda - \citet{Merm03}} database,
is far from complete, especially at the faint-end of the luminosity distribution (e.g. \citealt{Kharchenko05};
\citealt{DiskProp}). Besides, because of observational limitations associated with cluster/background contrast,
we actually observe a very small fraction of the OCs in the Galaxy (\citealt{DiskProp}). In this context,
derivation of astrophysical parameters of as yet unknown star clusters represents an important step to better
define their statistical properties.

Irrespective of the initial mass, star clusters evolve dynamically because of the combination of
internal and external processes. The main contributors to the internal processes are the mass
loss during stellar evolution, mass segregation and evaporation, while for the external ones are
tidal interactions with the disk and Galactic bulge, and collisions with giant molecular clouds
(GMCs). Consequently, the cluster structure changes significantly with age, to the point that
most (especially less-massive ones) end up completely dissolved in the Galactic stellar field
(e.g. \citealt{Lamers05}) or as poorly-populated remnants (e.g. \citealt{PB07}).

Probably reflecting the Galactocentric-dependence of most of the disruptive effects, the Galaxy presents
a spatial asymmetry in the age-distribution of OCs. Indeed, \citet{vdBMc80} noted that OCs older
than $\ga1$\,Gyr tended to be concentrated towards the anti-centre, a region with a low density of
GMCs. In this sense, the combined effect of tidal field and encounters with GMCs has been invoked
to explain the lack of old OCs in the solar neighbourhood (\citealt{Gieles06}, and references
therein). Near the Solar circle most OCs appear to dissolve on a time-scale shorter than $\sim1$\,Gyr
(\citealt{Bergond2001}; \citealt{DiskProp}), consistent with the disruption time-scales of
$\rm75\la\tdis(Myr)\la300$ for nearby clusters with mass in the range $10^2 - 10^3\ms$ (\citealt{Lamers05}).
In more central parts, interactions with the disk, the enhanced tidal pull of the Galactic bulge,
and the high frequency of collisions with GMCs tend to destroy the poorly-populated OCs on a time-scale
of a few $10^8$\,Myr (e.g. \citealt{Bergond2001}).

In general terms, the net effect of tidal interactions on a star cluster over long periods is to increase
the thermal energy. On average, member stars gain more kinetic energy after each event, leading to large-scale
mass segregation and an increase in the evaporation rate. Central tidal fields (at Galactocentric distances
$\dgc\la150$\,pc) can dissolve a massive star cluster in a very short time, $\tdis\approx50$\,Myr
(\citealt{Portegies02}). A discussion on the disruptive processes and associated time-scales can be found in
Bonatto \& Bica (2007a, and references therein).

Besides dynamical evolution-related effects, observational completeness also plays an important
r\^ole to explain the scarcity of open clusters in the inner Galaxy. High absorption and crowding in
fields dominated by disk and bulge stars are expected to significantly decrease completeness,
especially at the faint-end of the OC luminosity distribution. Indeed, \citet{DiskProp} found that
a large fraction of the intrinsically faint and/or distant OCs must be drowned in the field,
particularly in bulge/disk directions. Based on the spatial distribution of the Galactic OCs and
the related observational completeness, \citet{DiskProp} found that tidal disruption may be significant
for OCs located at distances $\ga1.4$\,kpc inside the Solar circle.

On the observational viewpoint, the arguments discussed above also reflect the importance of
deep, all-sky surveys to detect and characterise new star clusters, especially in central
directions. Such discoveries, in turn, can help constrain the Galactic tidal disruption
efficiency, improve statistics of the OC parameter space, and better define their age-distribution
function inside the Solar circle.

Recently, \citet{FSRcat} published a catalogue of 1021 star cluster candidates (hereafter FSR objects)
for $|b|<20^\circ$ and all Galactic longitudes, detected by means of an automated algorithm applied to 
the 2MASS\footnote{The Two Micron All Sky Survey, available at {\em
www.ipac.caltech.edu/2mass/releases/allsky/ }} database. They basically worked with stellar number-densities,
identifying regions with overdensities with respect to the surroundings. Each overdensity region was
classified according to a quality flag, '0' and '1' representing the most probable star clusters, while the
'5' and '6' flags may be related to field fluctuations. Nevertheless, we point out that CMDs are necessary
to try to distinguish star clusters from field density fluctuations.

The FSR catalogue has already produced two new GCs, FSR\,1735 (\citealt{FSR1735}) and  
FSR\,1767 (\citealt{FSR1767}), as well as the probable GC FSR\,584 (\citealt{FSR584}). Other prominent
clusters to visual inspection on the 2MASS Atlas, and later confirmed to be old open clusters, are
FSR\,1744, FSR\,89 and FSR\,31 (\citealt{OldOCs}), and FSR\,190 (\citealt{FSR190}).

Considering the above discoveries, it is fundamental to systematically explore the FSR catalogue guided
by the quality flags of the overdensities to look for star clusters. Our approach is based on 2MASS photometry,
on which we apply a field-star decontamination algorithm (\citealt{BB07}) that is essential to disentangle
physical from field CMD sequences. We also take into account properties of the stellar radial density profiles.

This paper is structured as follows. In Sect.~\ref{targets} we present fundamental data of the sample
targets. In Sect.~\ref{PhotPar} we present the 2MASS photometry and discuss the methods employed in the
CMD analyses, especially the field-star decontamination. In Sect.~\ref{RDPs} we analyse the stellar
radial density profiles and derive structural parameters of the confirmed star clusters. In Sect.~\ref{Disc}
we discuss the nature of the targets. Concluding remarks are given in Sect.~\ref{Conclu}.

\section{The FSR star cluster candidates}
\label{targets}

For the present study we selected all cluster candidates with quality flags '0' and '1' basically 
projected against the bulge ($|\ell|<60^\circ$), taken from both tables of probable and possible  
candidates (\citealt{FSRcat}).

Information on the resulting candidate sample are listed in Table~\ref{tab1}, where we also 
include the core and tidal radii measured by \citet{FSRcat} on the 2MASS \hh\ images by means of 
a \citet{King1962} profile fit. The FSR classification as probable or possible star cluster, as 
well as the quality flag are given. The targets in Table~\ref{tab1} are organised into three 
groups according to the nature implied by this work (Sect.~\ref{Disc}).

\begin{table}
\caption[]{General data on the FSR star cluster candidates}
\label{tab1}
\renewcommand{\tabcolsep}{0.76mm}
\renewcommand{\arraystretch}{1.2}
\begin{tabular}{lcccccccc}
\hline\hline
Target&$\alpha(2000)$&$\delta(2000)$&$\ell$&$b$&\rc&\rt&Class&Q\\
&(hms)&($^\circ\,\arcmin\,\arcsec$)&($^\circ$)&($^\circ$)&(\arcmin)&(\arcmin)\\
(1)&(2)&(3)&(4)&(5)&(6)&(7)&(8)&(9)\\
\hline
FSR\,70&19:30:02&$-$15:10:02& 23.4&$-15.3$&0.7&35.6&Poss&1 \\
FSR\,124&19:06:52&$+$13:15:21& 46.5&$+2.7$&1.6& 6.6&Prob&1 \\
FSR\,133&19:29:47&$+$15:34:28& 51.1&$-1.2$&3.8&11.6&Prob&1 \\
FSR\,1644&13:17:54&$-$67:03:28&305.5&$ -4.3$&2.0&10.0&Poss&1 \\
FSR\,1723&15:55:05&$-$46:00:51&333.0&$ +5.9$&1.1&11.5&Poss&1 \\
FSR\,1737&16:18:21&$-$40:14:35&340.1&$ +7.3$&1.7& 8.5&Poss&1 \\
\hline
FSR\,10&16:40:49&$-$16:01:09&  2.2&$+19.6$&1.4&13.6&Poss&0 \\
FSR\,98&18:47:31&$+$00:36:51& 33.0&$+1.2$&0.9&25.3&Prob&1 \\
FSR\,1740&17:49:16&$-$51:31:14&340.7&$-12.0$&1.1&14.8&Poss&1 \\
FSR\,1754&17:15:01&$-$39:06:07&348.0&$-0.3$&4.1& 8.2&Prob&1 \\
FSR\,1769&17:04:52&$-$31:02:16&353.3&$ +6.1$&1.0& 8.1&Prob&1 \\
\hline
FSR\,41&17:03:30&$-$08:51:13& 11.7&$+19.2$&2.5&39.2&Poss&1 \\
FSR\,91&17:38:21&$+$05:43:14& 29.7&$+18.9$&0.8& 3.4&Poss&1 \\
FSR\,114&20:09:09&$-$02:13:03& 40.0&$-18.3$&0.7&12.9&Poss&1 \\
FSR\,119&18:23:05&$+$15:49:12& 44.1&$+13.3$&1.5& 5.9&Poss&0 \\
FSR\,128&20:31:03&$+$04:42:51& 49.2&$-19.6$&0.8&12.0&Poss&1 \\
FSR\,1635&12:54:57&$-$43:29:24&303.6&$+19.4$&1.4&12.4&Poss&1 \\
FSR\,1647&13:45:49&$-$73:57:29&306.7&$-11.5$&2.0& 9.8&Poss&0 \\
FSR\,1685&14:57:22&$-$64:56:36&315.8&$ -5.3$&1.3&45.7&Poss&1 \\
FSR\,1695&14:33:38&$-$49:10:09&319.6&$+10.4$&0.8&17.9&Poss&1 \\
\hline
\end{tabular}
\begin{list}{Table Notes.}
\item Cols.~2-3: Central coordinates provided by \citet{FSRcat}. Cols.~4-5: Corresponding Galactic
coordinates. Cols.~6 and 7: Core and tidal radii derived by \citet{FSRcat} from King fits to the 
2MASS \hh\ images. Col.~8: FSR candidates have been classified by \citet{FSRcat} as probable star 
cluster (Prob) as possible clusters (Poss). Col.~9: FSR quality flag.
\end{list}
\end{table}

We verified that FSR\,1644 is the same OC as Harvard\,8 or Cr\,268 in early catalogues;
\citet{vdBH75} also included this object as BH\,145, and \citet{Lauberts82} as ESO\,96SC6.
FSR\,1723 corresponds to the optical OC ESO\,275SC1 (\citealt{Lauberts82}).

\section{Photometry and analytical tools}
\label{PhotPar}

In this section we briefly describe the photometry and outline the methods we apply in the CMD
analyses.

\subsection{2MASS photometry}
\label{2mass}

For each target we extracted \jj, \hh\ and \ks\ 2MASS photometry in a relatively wide circular field of
extraction radius \rx\ centred on the coordinates provided by \citet{FSRcat} (cols.~3 and 4 of Table~\ref{tab1})
using VizieR\footnote{\em vizier.u-strasbg.fr/viz-bin/VizieR?-source=II/246}. Such wide extraction areas
have been shown to provide
the required statistics, in terms of magnitude and colours, for a consistent field star decontamination
(Sect.~\ref{FSD}). They are essential as well to produce stellar radial density profiles with a high
contrast with respect to the background (Sect.~\ref{RDPs}). Our experience with OC analysis in dense
fields (e.g. \citealt{BB07}, and references therein) shows that both results can be reasonably well
achieved as long as no other populous cluster is present in the field, and differential absorption is
not prohibitive. In some cases the RDP resulting from the original FSR coordinates presented a dip
at the centre. For these we searched new coordinates that maximise the star-counts in the innermost
RDP bin. The optimised central coordinates are given in cols.~2 and 3 of Table~\ref{tab2}, while the
extraction radius is in col.~4.

As a photometric quality constraint, the 2MASS extractions were restricted to stars with errors in \jj,
\hh\ and \ks\ smaller than 0.25\,mag. A typical distribution of uncertainties as a function of magnitude,
for objects projected towards the central parts of the Galaxy, can be found in \citet{BB07}. About $75\% -
85\%$ of the stars have errors smaller than 0.06\,mag.

\subsection{Colour-magnitude diagrams}
\label{CMDs}

2MASS $\jj\times\jh$ and $\jj\times\jk$ CMDs extracted from a central ($R<2\arcmin$) region of FSR\,133
are presented in Fig.~\ref{fig1}. In this extraction that contains the bulk of the cluster stars
(Sect.~\ref{RDPs}), a cluster-like population ($0.6\la\jh\la1.0$ and $\jj\la16$) appears to mix with a redder
component ($\jh\ga1.0$). However, significant differences in terms of CMD densities are apparent with respect
to the equal-area comparison field (middle panels), extracted from a ring near $\rx$, which suggests some
contamination by disk and bulge stars. Such differences suggest the presence of a populous main sequence (MS)
for $13\la\jj\la16$, while the red component clearly presents an excess in the number of stars for $1.1\la\jh\la1.5$
and $13.3\la\jj\la14.7$ (top-left panel), which resembles a giant clump of an intermediate-age OC. Similar features
are present in the $\jj\times\jk$ CMD (top-right panel).

The observed CMDs of the remaining targets, extracted from central regions, are shown in the top panels of
Figs.~\ref{fig2} to \ref{fig5}, where for the sake of space, only the \jh\ CMDs are shown. Similarly to the
case of FSR\,133, essentially the same CMD features are present in both colours.

\subsection{Field-star decontamination}
\label{FSD}

\begin{figure}
\resizebox{\hsize}{!}{\includegraphics{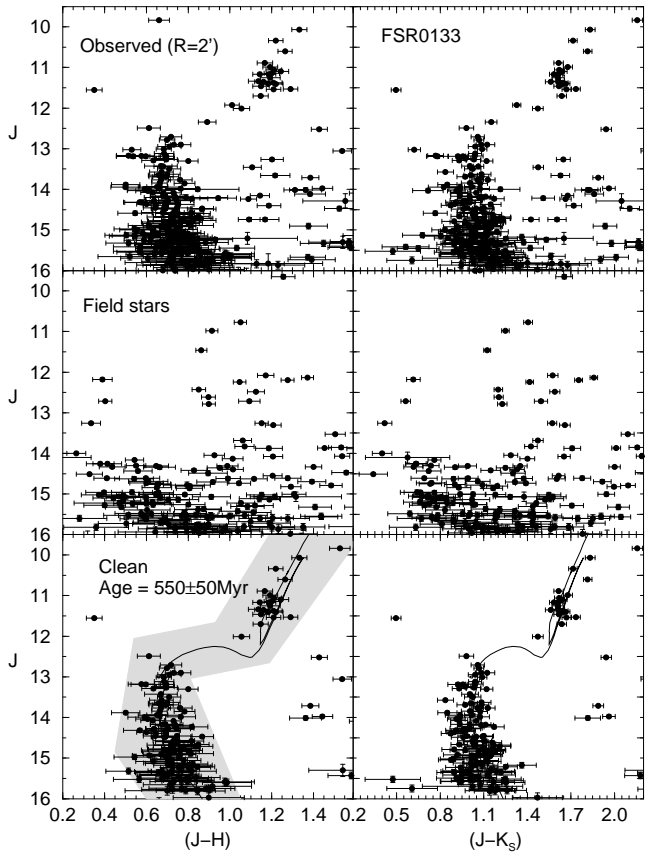}}
\caption{2MASS CMDs extracted from the $R<2\arcmin$ region of FSR\,133. Top panels: observed photometry
with the colours $\jj\times\jh$ (left) and $\jj\times\jk$ (right). Middle: equal-area comparison field.
Besides some contamination of disk and bulge stars, a populous MS and a conspicuous giant clump
($1.1\protect\la\jh\protect\la1.3$, $1.5\protect\la\jk\protect\la1.8$ and $10.8\protect\la\jj\protect\la11.8$)
show up. Bottom panels: decontaminated CMDs set with the 550\,Myr Padova isochrone (solid line). The
colour-magnitude filter used to isolate cluster MS/evolved stars is shown as a shaded region.}
\label{fig1}
\end{figure}

Features present in the central CMDs and those in the respective comparison field (top and middle panels
of Figs.~\ref{fig1} - \ref{fig5}), show that field stars contribute in varying proportions to the CMDs,
increasing in proportion especially for the cases projected close to the bulge, or at low Galactic latitudes.
In some cases, bulge contamination is the dominant feature, e.g. FSR\,1644, FSR\,1754 and FSR\,98. Nevertheless,
when compared to the equal-area offset-field extractions (middle panels), cluster-like sequences are suggested,
especially for FSR\,124, FSR\,1644 and FSR\,1723 (Fig.~\ref{fig2}). Thus, it is essential to assess the relative
densities of field stars and potential cluster sequences to determine the nature of the overdensities, whether
they are physical systems or field fluctuations.

To objectively quantify the field-star contamination in the CMDs we apply the statistical algorithm described 
in \cite{BB07}. It measures the relative number-densities of probable field and cluster stars in cubic CMD
cells whose axes correspond to the magnitude \jj\ and the colours \jh\ and \jk. These are the 2MASS colours 
that provide the maximum variance among CMD sequences for OCs of different ages (e.g. \citealt{TheoretIsoc}).
The algorithm {\em (i)} divides the full range of CMD magnitude and colours into a 3D grid, {\em (ii)} computes 
the expected number-density of field stars in each cell based on the number of comparison field stars with
similar magnitude and colours as those in the cell, and {\em (iii)} subtracts the expected number of field
stars from each cell. By construction, the algorithm is sensitive to local variations of field-star 
contamination with colour and magnitude (\citealt{BB07}). Typical cell dimensions are $\Delta\jj=0.5$, and
$\Delta\jh=\Delta\jk=0.25$, which are large enough to allow sufficient star-count statistics in individual
cells and small enough to preserve the morphology of different CMD evolutionary sequences. As comparison
field we use the region $\rl<R<\rx$ around the cluster centre to obtain representative background star-count
statistics, where \rl\ is the limiting radius (Sect.~\ref{RDPs}). We emphasise that the equal-area
extractions shown in the middle panels of Figs.~\ref{fig1} to \ref{fig5} serve only for visual comparison
purposes. Actually, the decontamination process is carried out with the large surrounding area as
described above. Further details on the algorithm, including discussions on subtraction efficiency 
and limitations, are given in \citet{BB07}.


Here we introduce the parameter \ns\ which, for a given extraction, corresponds to the ratio of the number of
stars in the decontaminated CMD with respect to the $\rm1\sigma$ Poisson fluctuation measured in the observed
CMD. By definition, CMDs of overdensities must have $\ns>1$. It is expected that CMDs of star
clusters have \ns\ significantly larger than 1. \ns\ values for the present sample are given in col.~5
of Table~\ref{tab2}.

By construction, \ns\ of a given extraction gives a measure of the statistical significance 
of the decontaminated number of stars integrated over all magnitudes. Thus, star clusters and 
field fluctuations should have $\ns>1$, although with different values (Sect.~\ref{Disc}). In 
this sense, it is also physically interesting to 
examine the dependence of \ns\ on magnitude. This analysis is presented in Table~\ref{tabSTAT}, 
which has been organised according to Sect.~\ref{Disc}. Because of the small number of bright
stars, we point out that this analysis should be considered basically for $\jj\ga10$. The spatial
regions considered here are those sampled by the CMDs shown in the top panels of Figs.~\ref{fig1}
to \ref{fig5}. The 
statistical significance of the number of probable member stars (\nc), which resulted from the 
decontamination algorithm, reaches the $2 - 3\sigma$ level (and even higher) with respect to the 
observed number of stars (\no), in most of the magnitude bins for the confirmed star clusters. 
This occurs especially in FSR\,133 and FSR\,1723. For the uncertain cases it decreases typically to 
the $\sim2\sigma$ level in most magnitude bins. In the above cases, the statistical significance 
per magnitude bin tends to systematically increase for fainter stars, as expected for a typical star 
cluster luminosity function (e.g. \citealt{DetAnalOCs}). However, the significance of the possible 
field fluctuations drops below the $\sim2\sigma$ level with no apparent dependence on magnitude, 
which is consistent with statistical fluctuations of a dense stellar field (Sect.~\ref{FieldFluc}).

To further test the statistical significance of the above results we compute the parameter
\sFS, which corresponds to the $\rm 1\,\sigma$ Poisson fluctuation around the mean of the star
counts measured in the 4 quadrants of the comparison field. By definition, \sFS\ is sensitive to
the spatial uniformity of the star counts in the comparison field. \sFS\ is computed for the same 
magnitude bins as before (Table~\ref{tabSTAT}). For a spatially uniform comparison field, \sFS\ 
should be very small. In this context, star clusters (and possible candidates) should have the
probable number of member stars (\nc) higher than $\sim3\,\sFS$. Indeed, this condition is fully
satisfied for the confirmed star clusters (Table~\ref{tabSTAT}) which, in some cases, reach the
level $\nc\sim10\,\sFS$. This ratio drops somewhat for the uncertain cases, reaching the minimum
for the possible field fluctuations. Similarly to above, we note that even for the latter cases, 
the number of probable member stars is higher than \sFS, which is consistent with the overdensity 
nature of these targets. 

The three statistical tests applied to the present sample, i.e. {\em (i)} the decontamination 
algorithm, {\em (ii)} the integrated and per magnitude \ns\ parameter, and {\em (iii)} the 
ratio of \nc\ to \sFS, produce consistent results.

\begin{table*}
\caption[]{Statistics of the field-star decontamination in magnitude bins}
\label{tabSTAT}
\renewcommand{\tabcolsep}{0.8mm}
\renewcommand{\arraystretch}{1.2}
\begin{tabular}{ccccccccccccccccccccccccc}
\hline\hline
\multicolumn{24}{c}{Confirmed star clusters}\\
\cline{1-24}
&\multicolumn{3}{c}{FSR\,70}&&\multicolumn{3}{c}{FSR\,124}&&\multicolumn{3}{c}{FSR\,133}&&
\multicolumn{3}{c}{FSR\,1644}&&\multicolumn{3}{c}{FSR\,1723}&&\multicolumn{3}{c}{FSR\,1737}\\
\cline{2-4}\cline{6-8}\cline{10-12}\cline{14-16}\cline{18-20}\cline{22-24}
$\Delta\jj$&\sFS&\no&\nc& &\sFS&\no&\nc& &\sFS&\no&\nc& &\sFS&\no&\nc& &\sFS&\no&\nc& &\sFS&\no&\nc \\
\cline{1-24}
  8-9&0.1&$1\pm1$&1  &&---&---&---&&---&---&---&&0.2&$1\pm1$&1  &&---&---&---&&0.1&$3\pm1.7$&3 \\
 9-10&---&---    &---&&---&---&---&&0.2&$1\pm1$&1  &&0.4&$4\pm2$&4&&0.1&$2\pm1.4$&2&&0.1&$1\pm1$&1 \\
10-11&0.4&$3\pm1.7$&3&&0.6&$3\pm1.7$&3&&0.3&$5\pm2.2$&5&&0.9&$6\pm2.5$&3&&0.2&$5\pm2.3$&5&&---&---&--- \\
11-12&0.3&$1\pm1$&1  &&0.8&$8\pm2.8$&7&&0.7&$18\pm4.2$&17&&2.3&$20\pm4.5$&10&&0.8&$10\pm3.2$&8&&0.2&$4\pm2$&4 \\
12-13&0.5&$6\pm2.4$&4&&2.6&$30\pm5.5$&22&&1.4&$13\pm3.6$&12&&4.0&$56\pm7.5$&26&&0.9&$20\pm4.5$&11&&0.7&$8\pm2.8$&5 \\
13-14&0.7&$12\pm3.5$&3&&5.1&$44\pm6.6$&25&&3.4&$40\pm6.3$&29&&10.4&$95\pm9.7$&36&&0.8&$47\pm6.8$&25&&1.0&$20\pm4.5$&9 \\
14-15&0.4&$38\pm6.2$&15&&9.3&$67\pm6.2$&24&&7.3&$83\pm9.1$&55&&19.4&$189\pm13.7$&40&&3.3&$87\pm9.3$&34&&1.8&$42\pm6.5$&20 \\
15-16&2.3&$61\pm7.8$&19&&1.7&$65\pm6.1$&13&&9.6&$113\pm10.6$&60&&19.7&$253\pm15.9$&21&&4.8&$149\pm12.2$&30&&2.3&$51\pm7.1$&17 \\
\hline
\multicolumn{24}{c}{Uncertain cases}\\
\cline{1-24}
&\multicolumn{3}{c}{FSR\,10}&&\multicolumn{3}{c}{FSR\,98}&&\multicolumn{3}{c}{FSR\,1740}&&
\multicolumn{3}{c}{FSR\,1754$^\dagger$}&&\multicolumn{3}{c}{FSR\,1754$^\ddagger$}&&\multicolumn{3}{c}{FSR\,1769}\\
\cline{2-4}\cline{6-8}\cline{10-12}\cline{14-16}\cline{18-20}\cline{22-24}
$\Delta\jj$&\sFS&\no&\nc& &\sFS&\no&\nc& &\sFS&\no&\nc& &\sFS&\no&\nc& &\sFS&\no&\nc& &\sFS&\no&\nc \\
\cline{1-24}
  8-9&---&---&---      &&---&---&---        &&---&---&---      &&0.4&$1\pm1$&1    &&---&---&---        &&---&---&---\\
 9-10&0.3&$1\pm1$&1    &&0.1&$2\pm1.4$&2    &&---&---&---      &&0.7&$4\pm2$&3    &&0.1&$1\pm1$&1      &&0.2&$2\pm1.4$&2 \\
10-11&0.4&$4\pm2$&4    &&1.1&$5\pm2.2$&5    &&0.1&$1\pm1$&1    &&1.6&$6\pm2.5$&5  &&0.5&$6\pm2.4$&5    &&0.3&$2\pm1.4$&1 \\
11-12&0.6&$5\pm2.2$&3  &&1.4&$17\pm4.1$&9   &&0.4&$1\pm1$&1    &&4.7&$20\pm4.5$&11&&2.5&$12\pm3.5$&6   &&0.6&$12\pm3.5$&7 \\
12-13&0.9&$12\pm3.5$&5 &&3.9&$38\pm6.2$&14  &&0.5&$7\pm2.6$&1  &&11.3&$41\pm6.4$&22&&4.5&$30\pm5.5$&14  &&1.6&$19\pm4.4$&11 \\
13-14&2.0&$21\pm4.6$&6 &&8.6&$78\pm8.8$&16  &&1.2&$28\pm5.3$&10&&19.6&$94\pm9.7$&35&&10.6&$84\pm9.2$&36  &&4.4&$46\pm6.8$&18 \\
14-15&3.4&$42\pm6.5$&14&&11.8&$191\pm13.8$&67&&0.4&$45\pm6.7$&13&&3.2&$7\pm2.6$&3  &&21.2&$173\pm13.2$&72&&9.0&$82\pm9.0$&18 \\
15-16&9.3&$69\pm8.3$&6 &&11.9&$257\pm16$&88  &&2.0&$88\pm9.4$&22&&---&---&---      &&7.7&$65\pm8.1$&43  &&1.8&$84\pm9.2$&25 \\
\hline

\multicolumn{24}{c}{Possible field fluctuations}\\
\cline{1-24}
&\multicolumn{3}{c}{FSR\,41}&&\multicolumn{3}{c}{FSR\,91}&&\multicolumn{3}{c}{FSR\,114}&&
\multicolumn{3}{c}{FSR\,119}&&\multicolumn{3}{c}{FSR\,128}&&\multicolumn{3}{c}{FSR\,1635}\\
\cline{2-4}\cline{6-8}\cline{10-12}\cline{14-16}\cline{18-20}\cline{22-24}
$\Delta\jj$&\sFS&\no&\nc& &\sFS&\no&\nc& &\sFS&\no&\nc& &\sFS&\no&\nc& &\sFS&\no&\nc& &\sFS&\no&\nc\\
\cline{1-24}
  8-9&---&---&---      &&---&---&---      &&---&---&---     &&---&---&---     &&---&---&---      &&---&---&---     \\
 9-10&0.1&$1\pm1$&1    &&---&---&---      &&---&---&---     &&0.1&$1\pm1$&1   &&0.1&$1\pm1$&1    &&---&---&---     \\
10-11&0.2&$1\pm1$&1    &&0.2&$2\pm1.4$&2  &&---&---&---     &&0.1&$3\pm1.7$&3 &&---&---&---      &&0.1&$1\pm1$&1   \\
11-12&0.2&$2\pm1.4$&2  &&0.1&$2\pm1.4$&2  &&---&---&---     &&0.1&$2\pm1.4$&2 &&0.1&$3\pm1.7$&3  &&---&---&---     \\
12-13&0.2&$4\pm2$&4    &&0.1&$3\pm1.7$&2  &&0.1&$2\pm1.4$&2 &&---&---&---     &&---&---&---      &&0.3&$3\pm1.7$&3 \\
13-14&0.2&$10\pm3.2$&7 &&0.2&$11\pm3.3$&8 &&0.1&$1\pm1$&1   &&0.6&$8\pm2.8$&7 &&0.2&$4\pm2$&3    &&0.2&$1\pm1$&1   \\
14-15&0.7&$17\pm4.1$&4 &&0.3&$23\pm4.8$&11&&0.1&$5\pm2.2$&5 &&0.7&$12\pm3.5$&7&&0.3&$6\pm2.4$&3  &&1.1&$10\pm3.2$&7 \\
15-16&1.1&$37\pm6.1$&10&&1.2&$34\pm5.8$&7 &&0.1&$10\pm3.2$&9&&0.9&$24\pm4.9$&5&&0.6&$18\pm4.2$&10&&0.4&$16\pm4$&8   \\
\hline
\end{tabular}

\begin{tabular}{cccccccccccc}
&\multicolumn{3}{c}{FSR\,1647}&&\multicolumn{3}{c}{FSR\,1685}&&\multicolumn{3}{c}{FSR\,1695}\\
\cline{2-4}\cline{6-8}\cline{10-12}
$\Delta\jj$&\sFS&\no&\nc& &\sFS&\no&\nc& &\sFS&\no&\nc\\
\cline{1-12}
  8-9&---&---&---     &&0.1&$3\pm1.7$&3 &&---&---&---\\
 9-10&---&---&---     &&0.1&$3\pm1.7$&3&&0.1&$1\pm1$&1\\
10-11&0.2&$1\pm1$&1   &&0.6&$4\pm2$&3&&---&---&--- \\
11-12&0.3&$5\pm2.2$&5 &&0.9&$5\pm2.2$&2&&0.2&$1\pm1$&1 \\
12-13&0.4&$2\pm1.4$& 1&&1.7&$28\pm5.3$&9&&0.4&$3\pm1.7$&2 \\
13-14&0.8&$3\pm1.7$&1 &&4.5&$55\pm7.4$&23&&0.9&$14\pm3.7$&8 \\
14-15&1.3&$13\pm3.6$&4&&6.1&$96\pm9.8$&26&&0.4&$23\pm4.8$&13\\
15-16&2.7&$24\pm4.9$&5&&6.9&$130\pm11.4$&10&&2.1&$47\pm6.8$&16 \\
\hline
\end{tabular}
\begin{list}{Table Notes.}
\item This table provides, for each magnitude bin ($\Delta\jj$), the $\rm 1\,\sigma$ Poisson fluctuation 
(\sFS) around the mean, with respect to the star counts measured in the 4 quadrants of the comparison field,  
the number of observed stars (\no) 
within the spatial region sampled in the CMDs shown in the top panels of Figs.~\ref{fig1} to \ref{fig5}, 
and the respective number of probable member stars (\nc) according to the decontamination algorithm. \nc\ 
can be compared to the $1\sigma$ Poisson fluctuation of \no. $(\dagger)$: FSR\,1754 as an IAC; $(\ddagger)$: 
FSR\,1754 as an old cluster.
\end{list}
\end{table*}

The resulting field star decontaminated CMDs of FSR\,133 are shown in the bottom panels of Fig.~\ref{fig1},
where for illustrative purposes CMDs in both colours, \jh\ and \jk, are shown. Bulge and disk contamination
have been properly taken into account, revealing conspicuous sequences, especially the giant clump and the MS,
typical of a relatively populous OC of $\rm age\approx550$\,Myr.

Five other objects resulted with cluster-like CMDs, FSR\,70, FSR\,124, FSR,1644, FSR\,1723, and FSR\,1737
(Fig.~\ref{fig2}). As expected, most of the disk and bulge components have been removed from their central 
CMDs. The decontaminated sequences of FSR\,124, FSR,1644 and FSR\,1723 are typical of OCs with ages in the 
range $\sim0.5-1.0$\,Gyr, while those of  FSR\,70 and FSR\,1737 suggest older ages. 

A second group composed by FSR\,10, FSR\,98, FSR\,1740, FSR\,1754 and  FSR\,1769 (Fig.~\ref{fig3}), 
end up with less-defined cluster-like sequences in the decontaminated CMDs. We cannot exclude the 
cluster possibility, but deeper observations are necessary.

Finally, the decontaminated CMDs of the remaining candidates (Figs.~\ref{fig4} and \ref{fig5}) do not
appear to present cluster-like sequences, instead the stellar distributions probably result from statistical
field fluctuations.

\subsection{Fundamental parameters}
\label{FundPar}

For the cases with a significant probability of being star clusters, we derive fundamental parameters
with solar-metallicity Padova isochrones (\citealt{Girardi02}) computed with the 2MASS \jj, \hh\ and
\ks\ filters\footnote{\em stev.oapd.inaf.it/$\sim$lgirardi/cgi-bin/cmd }. The 2MASS transmission filters
produced isochrones very similar to the Johnson-Kron-Cousins (e.g. \citealt{BesBret88}) ones, with
differences of at most 0.01 in \jh\ (\citealt{TheoretIsoc}).

The isochrone fit gives the age and the reddening \ejh, which converts to \ebv\ and \aV\ through the
transformations $A_J/A_V=0.276$, $A_H/A_V=0.176$, $A_{K_S}/A_V=0.118$, and $A_J=2.76\times\ejh$
(\citealt{DSB2002}), assuming a constant total-to-selective absorption ratio $R_V=3.1$. We also compute
the distance from the Sun (\ds) and the Galactocentric distance (\dgc), based on the recently derived value
of the Sun's distance to the Galactic centre $\Rgc=7.2$\,kpc (\cite{GCProp}). Age, \aV, \ds\ and \dgc\ are
given in cols.~7 to 10 of Table~\ref{tab2}, respectively. The isochrone fits to the probable star clusters
are shown in the bottom panels of Figs.~\ref{fig1} and \ref{fig2}.

\begin{table*}
\caption[]{2MASS fundamental parameters of the FSR star cluster candidates}
\label{tab2}
\renewcommand{\tabcolsep}{2.25mm}
\renewcommand{\arraystretch}{1.2}
\begin{tabular}{lccccccccc}
\hline\hline
Target&$\alpha(2000)$&$\delta(2000)$&\rx&\ns&$\rm Q_{RDP}$&Age&\aV&\ds&\dgc\\
&(hms)&($^\circ\,\arcmin\,\arcsec$)&(\arcmin)&&&(Gyr)&(mag)&(kpc)&(kpc)\\
(1)&(2)&(3)&(4)&(5)&(6)&(7)&(8)&(9)&(10)\\
\hline
\multicolumn{10}{c}{Confirmed star clusters}\\
\hline
FSR\,70&($\ddagger$)& ($\ddagger$)&30&5.5&K&$\approx5$&$0.8\pm0.1$&$2.3\pm0.2$&$5.3\pm0.3$\\
FSR\,124&($\ddagger$)& ($\ddagger$)&240&5.5&K&$1.0\pm0.2$&$3.4\pm0.2$&$2.6\pm0.1$&$5.7\pm0.2$\\
FSR\,133&19:29:48.5&$+$15:33:36&60&9.7&K&$0.6\pm0.1$&$6.3\pm0.2$&$1.9\pm0.1$&$6.2\pm0.2$\\
Harvard\,8,Cr\,268,FSR\,1644&13:18:2.9&$-$67:04:33.6&40&4.9&K&$0.6\pm0.1$&$0.7\pm0.1$&$1.9\pm0.1$&$6.3\pm0.2$
     \\
ESO\,275SC1,FSR\,1723&($\ddagger$)& ($\ddagger$)&40&5.9&K&$0.8\pm0.1$&$0.1\pm0.1$&$1.3\pm0.1$&$6.1\pm0.2$\\
FSR\,1737&($\ddagger$)& ($\ddagger$)&40&5.6&K&$\ga5$&$1.8\pm0.1$&$2.8\pm0.1$&$4.7\pm0.2$\\
\hline
\multicolumn{10}{c}{Uncertain cases: deeper photometry necessary}\\
\hline
FSR\,10&($\ddagger$)& ($\ddagger$)&50&4.5&M&IAC?&---&---&---\\
FSR\,98&18:47:36&$+$00:35:45.6&40&7.0&M&Old cluster?\\
FSR\,1740&17:49:17.8&$-$51:31:55.2&40&5.3&M&Old cluster?&---&---&---\\
FSR\,1754&17:15:3.4&$-$39:05:45.6&60&5.5& &IAC?&---&---&---\\
FSR\,1754&17:15:3.4&$-$39:05:45.6&60&9.8& &Old cluster?&---&---&---\\
FSR\,1769&17:04:41.3&$-$31:00:43.2&40&4.9&M&Old cluster?\\
\hline
\multicolumn{10}{c}{Possible field fluctuations}\\
\hline
FSR\,41&($\ddagger$)& ($\ddagger$)&40&3.4&FF&---&---&---&---\\
FSR\,91&($\ddagger$)& ($\ddagger$)&40&3.3&FF&---&---&---&---\\
FSR\,114&($\ddagger$)& ($\ddagger$)&20&3.5&M&---&---&---&---\\
FSR\,119&($\ddagger$)& ($\ddagger$)&30&3.4&FF&---&---&---&---\\
FSR\,128&20:31:10.1&$+$04:45:7.2&40&3.4&M&---&---&---&---\\
FSR\,1635&($\ddagger$)& ($\ddagger$)&20&3.8&FF&---&---&---&---\\
FSR\,1647&13:45:48&$-$73:57:28.8&30&2.6&M&---&---&---&---\\
FSR\,1685&14:57:14.9&$-$64:57:21.6&40&3.5&FF&---&---&---&---\\
FSR\,1695&($\ddagger$)& ($\ddagger$)&20&4.3&M&---&---&---&---\\
\hline
\end{tabular}
\begin{list}{Table Notes.}
\item Cols.~2 and 3: Optimised central coordinates (Sect.~\ref{2mass}); $(\ddagger)$ indicates 
same central coordinates as in \citet{FSRcat}. Col.~4: 2MASS extraction radius. Col.~5: Ratio of 
the decontaminated star-counts to the $\rm1\sigma$ fluctuation level of the observed photometry.
Col.~6: RDP quality flag, with K: RDP follows a King profile, M: medium quality and FF:
possibly a field fluctuation. IAC in col.~7 means intermediate-age cluster. Col.~8: 
$\rm\aV=3.1\,\ebv$. Col.~10: \dgc\ calculated using $\Rgc=7.2$\,kpc (\citealt{GCProp}) as the 
distance of the Sun to the Galactic centre.
\end{list}
\end{table*}

\begin{figure}
\caption{Same as Fig.~\ref{fig1} for the $\jj\times\jh$ CMDs of the central regions
of the remaining confirmed star clusters.}
\label{fig2}
\end{figure}

\begin{figure}
\caption{Same as Fig.~\ref{fig1} for the $\jj\times\jh$ CMDs of the central regions
of the uncertain cases. Two sequences are conspicuous in the decontaminated CMD of FSR\,1754,
a blue one that suggests an IAC and a red one that suggests an older cluster. }
\label{fig3}
\end{figure}

\begin{figure}
\resizebox{\hsize}{!}{\includegraphics{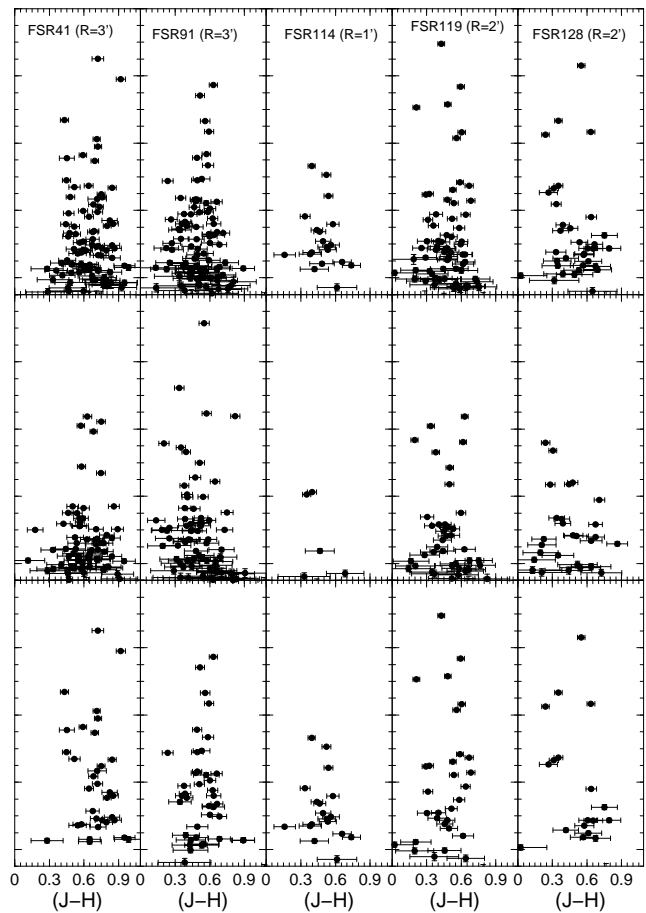}}
\caption{Same as Fig.~\ref{fig1} for the $\jj\times\jh$ CMDs of part of the possible field
fluctuations.}
\label{fig4}
\end{figure}

\begin{figure}
\resizebox{\hsize}{!}{\includegraphics{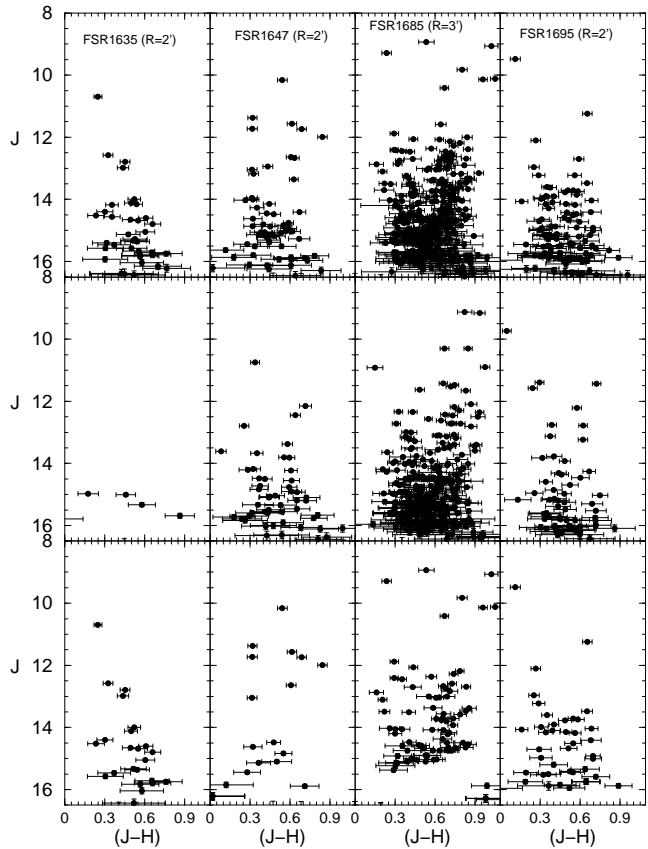}}
\caption{Same as Fig.~\ref{fig4} for the remaining possible field fluctuations.}
\label{fig5}
\end{figure}

\subsection{Colour-magnitude filters}
\label{CMF}

Colour-magnitude filters are used to exclude stars with colours compatible with those of the
foreground/background field. They are wide enough to accommodate cluster MS and evolved star
colour distributions, allowing for the $1\sigma$ photometric uncertainties. Colour-magnitude
filter widths should also account for formation or dynamical evolution-related effects, such
as enhanced fractions of binaries (and other multiple systems) towards the central parts of
clusters, since such systems tend to widen the MS (e.g. \citealt{BB07}; \citealt{N188};
\citealt{HT98}; \citealt{Kerber02}).

However, residual field stars with colours similar to those of the cluster are
expected to remain inside the colour-magnitude filter region. They affect the intrinsic stellar
radial distribution profile in a degree that depends on the relative densities of field and cluster
stars. The contribution of the residual contamination to the observed RDP is statistically taken
into account by means of the comparison field. In practical terms, the use of colour-magnitude
filters in cluster sequences enhances the contrast of the RDP with respect to the background level,
especially for objects in dense fields (e.g. \citealt{BB07}; see Sect.~\ref{RDPs}).

\section{Stellar radial density profiles}
\label{RDPs}

As another clue to the nature of the overdensities, we investigate properties of the stellar radial
density profiles. Star clusters usually have RDPs that follow some well-defined analytical profile.
The most often used are the single-mass, modified isothermal sphere of \citet{King66}, the modified
isothermal sphere of \citet{Wilson75}, and the power-law with a core of \citet{EFF87}. Each function
is characterised by different parameters that are somehow related to cluster structure. However,
considering that error bars in the present RDPs are significant (see below), and that our goal here
is basically to determine the nature of the targets, we decided for the
analytical profile $\sigma(R)=\sigma_{bg}+\sigma_0/(1+(R/R_C)^2)$, where $\sigma_{bg}$ is
the residual background density, $\sigma_0$ is the central density of stars, and \rc\ is the core
radius. This function is similar to that introduced by \cite{King1962} to describe the surface
brightness profiles in the central parts of globular clusters.

In all cases we build the stellar RDPs with colour-magnitude filtered photometry (Sect.~\ref{CMF}).
To avoid oversampling near the centre and undersampling at large radii, RDPs are built by counting
stars in rings of increasing width with distance to the centre. The number and width of the rings are
adjusted to produce RDPs with adequate spatial resolution and as small as possible $1\sigma$ Poisson
errors. The residual background level of each RDP corresponds to the average number of colour-magnitude
filtered stars measured in the comparison field. The $R$ coordinate (and respective uncertainty) of each
ring corresponds to the average position and standard deviation of the stars inside the ring.

\begin{figure}
\resizebox{\hsize}{!}{\includegraphics{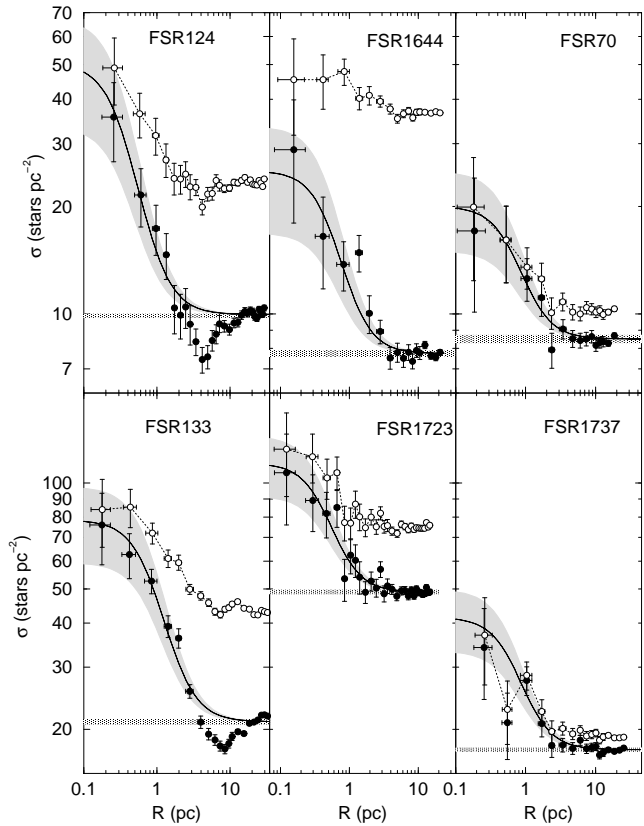}}
\caption{Stellar RDPs (filled circles) of the confirmed star clusters built with colour-magnitude photometry.
Solid line: best-fit King profile. Horizontal shaded region: offset field stellar background level. Gray
regions: $1\sigma$ King fit uncertainty. RDPs built with the observed photometry (empty circles) are shown
for comparison. Absolute scale is used.}
\label{fig6}
\end{figure}

The resulting radial profiles of the 6 confirmed star clusters are given in Fig.~\ref{fig6}. Besides
the RDPs resulting from the colour-magnitude filters, we also show, for illustrative purposes, those
produced with the observed (raw) photometry. Minimisation of non-cluster stars by the colour-magnitude 
filter resulted in RDPs with a significantly higher contrast with the background, especially for FSR1644, 
FSR124 and FSR\,133. As expected for star clusters, the adopted King-like function describes well the RDPs
throughout the full radii range, within uncertainties. $\sigma_0$ and the core radius (\rc) are derived 
from the RDP fit, while $\sigma_{bg}$ is measured in the respective comparison field. These values are 
given in Table~\ref{tab4}, and the best-fit solutions are superimposed on the colour-magnitude filtered 
RDPs (Fig.~\ref{fig6}). Because of the 2MASS photometric limit, which in most cases corresponds to a cutoff 
for stars brighter than $\jj\approx16$, $\sigma_0$ should be taken as a lower limit to the actual central
number-density.

The intrinsic contrast of a cluster RDP with the background which, in turn, is related to the
difficulty of detection, can be quantified by the density contrast parameter
$\delta_c=1+\sigma_0/\sigma_{bg}$ (col.~5 of Table~\ref{tab4}). Interestingly, the objects projected
not close to the Galactic centre, FSR\,124, FSR\,133 and FSR\,1644, with $|\Delta\ell|\ga45^\circ$, 
have $\delta_c\ga3.3$, while the more central ones have $\delta_c\approx2$. As a caveat we note that
since $\delta_c$ is measured in colour-magnitude filtered RDPs, it does not necessarily correspond to
the visual contrast produced in optical/IR images. The values of $\delta_c$ quoted in Table~\ref{tab4}
are larger than the observed ones, as can be clearly seen in the observed RDPs (Fig.~\ref{fig6}).

\begin{table*}
\caption[]{Structural parameters measured in the RDPs built with colour-magnitude filtered photometry}
\label{tab4}
\renewcommand{\tabcolsep}{5.0mm}
\renewcommand{\arraystretch}{1.3}
\begin{tabular}{lcccccccc}
\hline\hline
&&&\multicolumn{5}{c}{RDP}\\
\cline{4-8}
Cluster&$1\arcmin$&&$\sigma_{bg}$&$\sigma_0$&$\delta_c$&\rc&\rl \\
       &(pc)&&$\rm(stars\,pc^{-2})$&$\rm(stars\,pc^{-2})$&&(pc)&(pc)\\
(1)&(2)&&(3)&(4)&(5)&(6)&(7)\\
\hline
FSR\,70  &0.658&&$9.6\pm0.1$&$9.9\pm4.4$&$2.0\pm0.5$&$0.7\pm0.2$&$3.3\pm0.6$\\
FSR\,124  &0.749&&$10.0\pm0.1$&$40.4\pm16.9$&$5.1\pm1.7$&$0.4\pm0.1$ &$4.0\pm1.0$\\
FSR\,133  &0.561&&$21.0\pm0.1$&$57.4\pm19.0$&$3.7\pm0.9$&$0.9\pm0.2$ &$4.4\pm0.4$\\
Harvard\,8,Cr\,268$^{\dagger}$&0.554&&$7.7\pm0.1$&$17.4\pm8.3$&$3.3\pm1.0$&$0.6\pm0.2$&$3.7\pm0.3$\\
ESO\,275SC1$^{\ddagger}$&0.365&&$49.0\pm0.3$&$64.6\pm21.9$&$2.3\pm0.4$&$0.5\pm0.1$ &$3.6\pm0.6$\\
FSR\,1737  &0.800&&$12.2\pm0.1$&$16.6\pm5.6$&$2.4\pm0.2$&$0.8\pm0.2$ &$5.0\pm0.3$\\
\hline
\end{tabular}
\begin{list}{Table Notes.}
\item $(\dagger)$: FSR\,1644; $(\ddagger)$: FSR\,1723.
Col.~2: arcmin to parsec scale. To minimise degrees of freedom in RDP fits with the King-like
profile (see text), $\sigma_{bg}$ was kept fixed (measured in the respective comparison fields) while
$\sigma_0$ and \rc\ were allowed to vary. Col.~5: cluster/background density contrast
($\delta_c=1+\sigma_0/\sigma_{bg}$), measured in colour-magnitude filtered RDPs.
\end{list}
\end{table*}

\begin{figure}
\resizebox{\hsize}{!}{\includegraphics{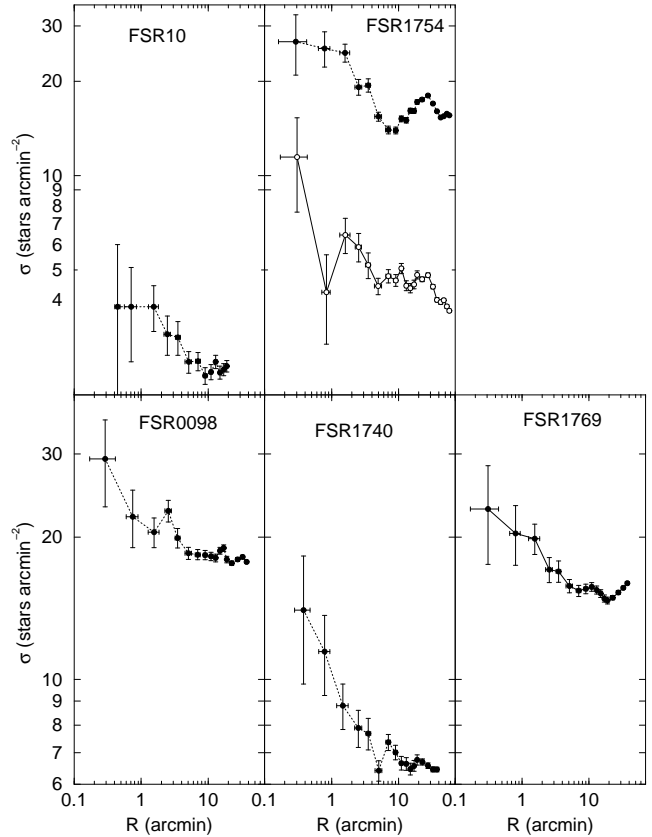}}
\caption{RDPs of the uncertain cases, in angular units. FSR\,1754 has two RDPs corresponding
to the red (black circles) and blue (empty circles) sequences seen in the decontaminated CMD
(Fig.~\ref{fig3}). The "bump" at $\rm R\sim30\arcmin$ in the RDP of FSR\,1754 is produced by
the OC NGC\,6318.}
\label{fig7}
\end{figure}

\begin{figure}
\resizebox{\hsize}{!}{\includegraphics{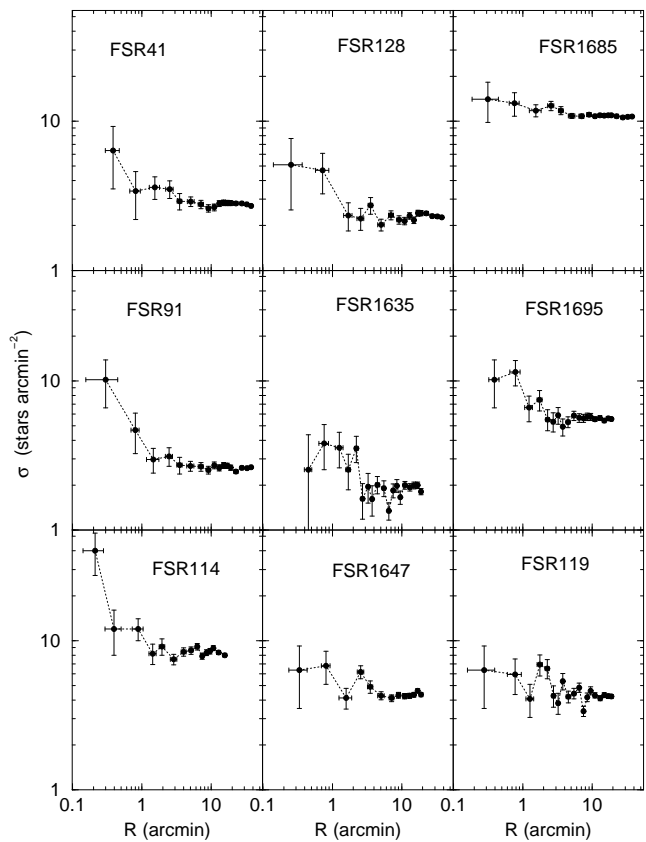}}
\caption{RDPs of the possible field fluctuations, in angular units. They are narrower than
the RDPs of the cluster-like (Fig.~\ref{fig6} and uncertain cases.}
\label{fig8}
\end{figure}

We also provide in col.~7 of Table~\ref{tab4} the cluster limiting radius and uncertainty, which are
estimated by comparing the RDP (taking into account fluctuations) with the background level. \rl\
corresponds to the distance from the cluster centre where RDP and background become statistically
indistinguishable. For practical purposes, most of the cluster stars are contained within $\rl$. The
limiting radius should not be mistaken for the tidal radius; the latter values are usually derived from
King (or other analytical functions) fits to RDPs, which depend on wide surrounding fields and as small
as possible Poisson errors (e.g. \citealt{BB07}). In contrast, \rl\ comes from a visual comparison
of the RDP and background level.

The empirical determination of a cluster-limiting radius depends on the relative levels of RDP and
background (and respective fluctuations). Thus, dynamical evolution may indirectly affect the
measurement of the limiting radius. Since mass segregation preferentially drives low-mass stars to the
outer parts of clusters, the cluster/background contrast in these regions tends to lower as clusters age.
As an observational consequence, smaller values of limiting radii should be measured, especially for
clusters in dense fields. However, simulations of King-like OCs (\citealt{BB07}) show that, provided not
exceedingly high, background levels may produce limiting radii underestimated by about 10--20\%. The core
radius, on the other hand, is almost insensitive to background levels (\citealt{BB07}). This occurs because
\rc\ results from fitting the King-like profile to a distribution of RDP points, which minimises background
effects.

The RDPs of the cases with uncertain CMD morphology are shown in Fig.~\ref{fig7}. Except for FSR\,10,
which suffers from low-number statistics, the remaining RDPs suggest the presence of a star cluster.

Finally, in Fig.~\ref{fig8} we show the RDPs of the remaining targets. A narrow excess in the stellar 
density profile near the centre is present in all cases, but they are quite different from a King-like
profile (e.g. Fig.~\ref{fig6}).

\section{Discussion}
\label{Disc}

Following the photometric (Sect.~\ref{PhotPar}) and RDP (Sect.~\ref{RDPs}) analyses, the 20
FSR overdensity/star cluster candidates dealt with in this paper can be split into the three 
distinct groups discussed below.

\subsection{Star clusters}
\label{ConfClus}

Objects in the first group have well-defined decontaminated CMD sequences (Figs.~\ref{fig1} and
\ref{fig2}) with relatively high values of the parameter \ns, both considering magnitude
bins (Table~\ref{tabSTAT}) and the integrated one (Table~\ref{tab2}), as well as King-like
RDPs (Fig.~\ref{fig6}). In most cases, the statistical significance of the decontaminated
number of probable member stars, in individual magnitude bins, is $\ga3\sigma$ with respect to 
fluctuations in the observed number of stars. Astrophysical parameters (age, distance, reddening, 
core and limiting radii) 
could be measured for these clusters. They are FSR\,70, FSR\,124, FSR\,133, FSR\,1644, FSR\,1723 and 
FSR\,1737. The average value of \ns\ is $\langle\ns\rangle=6.2\pm1.7$.

{\tt FSR\,70:} The decontaminated CMD ($\ns=5.5$) is typical of an old cluster. In Fig.~\ref{fig2}
we tentatively applied the 5\,Gyr isochrone, which resulted in a distance from the Sun of
$\ds\approx2.3$\,kpc, and the Galactocentric distance $\dgc\approx5.3$\,kpc. The RDP, with a density
contrast $\delta_c\approx2$, produced the structural parameters $\rc\approx0.7$\,pc and $\rl\approx3.3$\,pc.
Within uncertainties, the present \rc\ value agrees with that computed by \citet{FSRcat} (Table~\ref{tab1}).

{\tt FSR\,124:} The presence of an intermediate-age OC was already suggested by the observed CMD. 
With $\ns=5.5$ in the decontaminated CMD, we derived the age $\approx1$\,Gyr, $\ds\approx2.6$\,kpc
and $\dgc\approx5.7$\,kpc. From the highly contrasted RDP ($\delta_c\approx5$) we derived 
$\rc\approx0.4$\,pc and $\rl\approx4.0$\,pc. In this case our value of \rc\ is $\sim1/3$ that
in \citet{FSRcat}.

{\tt FSR\,133:} This OC presents the highest reddening value ($\aV\approx6$) among the present sample.
It appears to be the most populous as well, with the decontaminated $\ns\approx10$. We found the age
$\approx550$\,Myr, $\ds\approx1.9$\,kpc and $\dgc\approx6.2$\,kpc. From the RDP ($\delta_c\approx4$) we 
derived $\rc\approx0.9$\,pc ($\sim1/2$ that in \citealt{FSRcat}) and $\rl\approx4.4$\,pc. 

{\tt Harvard\,8, Cr\,268 = FSR\,1644:} The decontamination ($\ns=4.9$) was essential to uncover this OC
with the age $\approx550$\,Myr, at $\ds\approx1.9$\,kpc and $\dgc\approx6.3$\,kpc. From the RDP 
($\delta_c\approx3.3$) we derived $\rc\approx0.6$\,pc ($\sim1/2$ that in \citealt{FSRcat}) and
$\rl\approx3.7$\,pc. WEBDA provides for this
optical cluster under the designation Cr\,268, $\ebv=0.31$, $\ds=1.96$\,kpc and the age 0.57\,Gyr, in 
excellent agreement with the present work (Table~\ref{tab2}).

{\tt ESO\,275SC1 = FSR\,1723:} An OC of age $\approx0.8$\,Gyr, at $\ds\approx1.3$\,kpc and $\dgc\approx6.1$\,kpc,
clearly stands out both in the observed and decontaminated ($\ns=5.9$) CMDs, which presents the lowest
reddening ($\aV\approx0.1$) among the sample. The King-like RDP ($\delta_c\approx2.3$) is characterised by
$\rc\approx0.5$\,pc (similar to that in \citealt{FSRcat}) and $\rl\approx3.6$\,pc.

{\tt FSR\,1737:} Another case of an old OC whose decontaminated CMD ($\ns=5.6$) suggests an age of
5\,Gyr, or older. In the case of 5\,Gyr, we estimated $\ds\approx2.8$\,kpc and $\dgc\approx4.7$\,kpc.
Its RDP ($\delta_c\approx2.4$) implies $\rc\approx0.8$\,pc ($\sim1/2$ that in \citealt{FSRcat})
and $\rl\approx5.0$\,pc.

\subsection{Uncertain cases}
\label{DubCases}

In general, targets of the second group have less defined decontaminated CMD sequences than
those in the first group, which is consistent with the lower-level of the \ns\ parameter in magnitude 
bins, which reaches a statistical significance of $\sim2\sigma$; however, their integrated \ns\ 
are, on average, of the same order ($\langle\ns\rangle=6.1\pm1.8$). 
They are FSR\,10, FSR\,98, FSR\,1740, FSR\,1769 and FSR\,1754. Cluster sequences are suggested by 
the decontaminated CMDs (Fig.~\ref{fig3}), e.g. giant clump and the top of the MS. RDPs of the objects
in this group (Fig.~\ref{fig7}) also suggest star clusters, although the large error bars of FSR\,10
reflect the low-number statistics.

FSR\,1754 is an interesting case whose decontaminated CMD presents two sequences, a blue one with
$\ns=5.5$ and a more populous red one with $\ns=9.8$. The former may be from an intermediate-age
cluster (IAC), while the latter might correspond to an old cluster. RDPs extracted from both sequences
separately (Fig.~\ref{fig7}) also suggest star clusters. We point out that the field of FSR\,1754
contains the OC NGC\,6318, at $\approx26\arcmin$ from the centre (WEBDA), which can be seen in the RDP
of FSR\,1754 as a "bump" on the wing (Fig.~\ref{fig7}).

Decontaminated CMDs and RDPs taken together suggest that the above objects might be old clusters
which require deeper observations. FSR\,10, on the other hand, may be an IAC. Deeper photometry
is essential in most cases, especially for old OCs for which the TO is close to the 2MASS 
limiting magnitude. In this context, we would recommend also that the same applies to FSR\,70 and
FSR\,1737 (Sect.~\ref{ConfClus}), for a better definition of the TO region and, consequently, the
age and distance from the Sun.

\subsection{Possible field fluctuations}
\label{FieldFluc}

Decontaminated CMDs of this group have \ns-values significantly lower than those of the star 
clusters (Sect.~\ref{ConfClus}) and uncertain cases (Sect.~\ref{DubCases}). Indeed, the average 
integrated \ns\ is $\langle\ns\rangle=3.5\pm0.5$, while the statistical significance of the probable
member stars in individual magnitude bins is below the $2\sigma$ level. The
fact that they have $\ns\sim3$ is consistent with the method employed by \citet{FSRcat} to detect
overdensities. However, in most cases the RDP excess is very narrow, restricted to the first bins, 
quite different from a King-like profile (e.g. Fig.~\ref{fig6}).

The third group has essentially featureless (observed and decontaminated) CMDs, and RDPs with
important deviations from cluster-like profiles. They appear to be $\sim3\sigma$ fluctuations
of the dense stellar field over which these objects are projected.

\subsection{Relations among astrophysical parameters}
\label{Relat}

\begin{figure}
\resizebox{\hsize}{!}{\includegraphics{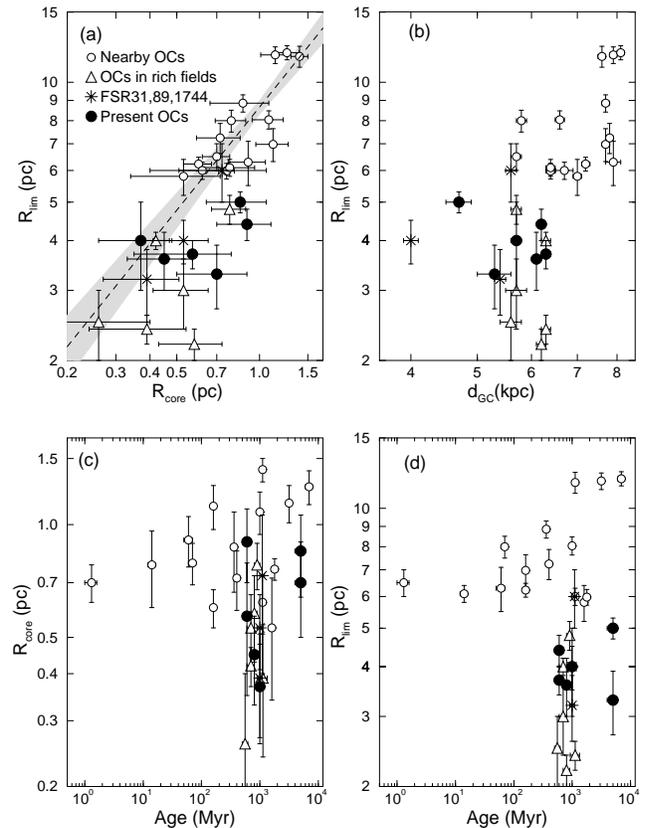}}
\caption{Relations involving astrophysical parameters of OCs. Empty circles: nearby OCs, including
two young ones. Triangles: OCs projected on dense fields towards the centre. Stars: the similar
OCs FSR\,31, FSR\,89 and FSR\,1744. Black circles: the OCs dealt with in this work.}
\label{fig9}
\end{figure}

To put the present FSR OCs into perspective we compare in Fig.~\ref{fig9} their astrophysical 
parameters with those measured in OCs in different environments. We consider {\em (i)} a sample 
of bright nearby OCs (\citealt{DetAnalOCs}), including the two young OCs NGC\,6611 (\citealt{N6611}) 
and NGC\,4755 (\citealt{N4755}), {\em (ii)} OCs projected against the central parts of the Galaxy 
(\citealt{BB07}), and {\em (iii)} the recently analysed OCs FSR\,1744, FSR\,89 and FSR\,31 
(\citealt{OldOCs}), which are similarly projected against the central parts of the Galaxy as the
present FSR cluster sample {\em (iv}). OCs in sample {\em (i)} have ages in the range 
$\rm70\,Myr\la age\la7\,Gyr$ and Galactocentric distances in the range $\rm5.8\la\dgc(kpc)\la8.1$.
NGC\,6611 has $\rm age\approx1.3$\,Myr and $\dgc=5.5$\,kpc, and NGC\,4755 has $\rm age\approx14$\,Myr 
and $\dgc=6.4$\,kpc. Sample {\em (ii)} OCs are characterised by $\rm600\,Myr\la age\la1.3\,Gyr$ and
$\rm5.6\la\dgc(kpc)\la6.3$. FSR\,1744, FSR\,89 and FSR\,31 are Gyr-class OCs at $\rm4.0\la\dgc(kpc)\la5.6$.

Core and limiting radii of the OCs in samples {\em (i)} and {\em (ii)} are almost linearly related by
$\rl=(8.9\pm0.3)\times R_{\rm core}^{(1.0\pm0.1)}$ (panel (a)), which suggests a similar scaling in both 
kinds of radii, in the sense that on average, larger clusters tend to have larger cores, at least for
$\rm 0.5\la\rc(pc)\la1.5$ and $\rm 5\la\rl(pc)\la15$. Linear relations between OC core and limiting radii were
also found by \citet{Nilakshi02}, \citet{Sharma06}, and \citet{MacNie07}. However, about $2/3$ of the OCs 
in samples {\em (iii)} and {\em (iv)} do not follow that relation, which suggests that they are either 
intrinsically small or have suffered important evaporation effects (see below). The core and 
limiting radii of FSR\,124 and FSR\,1723 are consistent with the relation at the $1\sigma$ level.

A dependence of OC size on Galactocentric distance is implied by panel (b), as previously suggested by 
\citet{Lynga82} and \citet{Tad2002}. In this context, the limiting radii of the present FSR OCs are 
roughly consistent with their positions in the Galaxy, especially FSR\,1737, the innermost OC of the 
sample. Since core and limiting radii appear to be linearly related (panel a), a similar conclusion 
applies to the core radius. Part of this relation may be primordial, in the sense that the higher density
of molecular gas in central Galactic regions may have produced clusters with smaller core radii,
as suggested by \citet{vdBMP91} to explain the increase of GC radii with Galactocentric
distance. In addition, there is the possibility that the core size may also be a function of 
the binary fraction and its evolution with age, so that loss of stars may not be the only
process to determine sizes. 

Core and limiting radii are compared with cluster age in panels (c) and (d), respectively. This
relationship is intimately related to cluster survival/dissociation rates. Both kinds of radii
present a similar dependence on age, in which part of the clusters expand with time, while some
seem to shrink. The bifurcation occurs at $\rm age\sim1$\,Gyr. Except for FSR\,133 (and perhaps,
FSR\,1737), the remaining FSR OCs have core radii typical of the small OCs in the lower branch;
the limiting radii, on the other hand, locate in the lower branch.

With respect to the astrophysical parameters discussed above, the present FSR star clusters can 
be taken as similar objects as FSR\,1744, FSR\,89 and FSR\,31 (\citealt{OldOCs}). In that study we
interpreted the relatively small radii of the latter OCs as resulting from the enhanced dynamical
evolution combined to low-contrast. Effects such as the tidal pull of the Galactic bulge, frequency 
of collisions with giant molecular clouds and spiral arms, low-mass star evaporation and ejection,
which are more important in the inner Galaxy, tend to accelerate the dynamical evolution, especially 
of low-mass star clusters (\citealt{OldOCs}, and references therein). As a result, the mass of the 
OCs decreases with time. 

One consequence of the mass segregation associated to the dynamical evolution is the large-scale
transfer of low-mass stars towards the external parts, which reduces the surface brightness at 
large radii. When projected against the central parts of the Galaxy, such star clusters (as
well as the poorly-populated ones) suffer from low-contrast effects, especially in the external
parts. \citet{BB07} found that low contrast may underestimate the limiting radii of centrally
projected OCs by about $10 - 20\%$. The core radii, on the other hand, are not affected. Thus,
the small sizes of the present FSR clusters derived here appear to be related to 
dynamical effects.

\begin{figure}
\resizebox{\hsize}{!}{\includegraphics{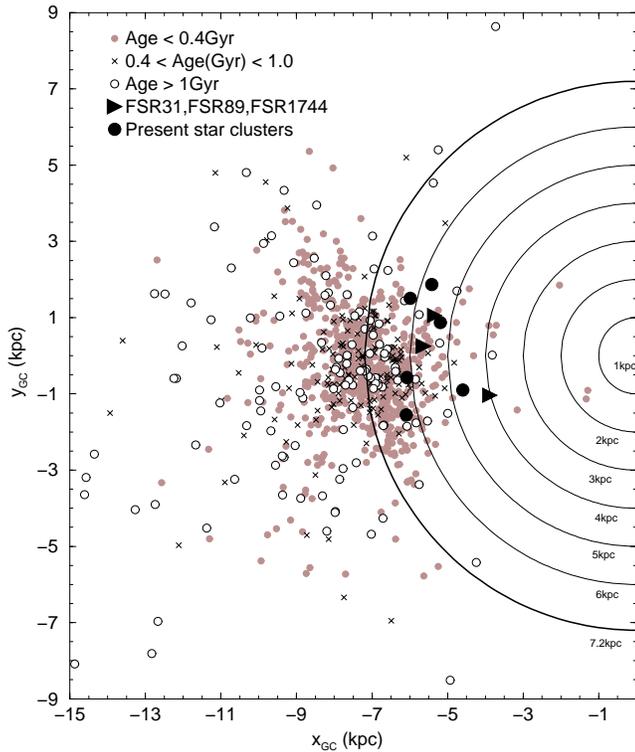}}
\caption{Spatial distribution of the present star clusters (black circles) compared to the
WEBDA OCs with ages younger than 0.4\,Gyr (gray circles), $\rm0.4<age<1.0$\,Gyr ('x'), and
older than 1\,Gyr (empty circles). For comparison we also show the position of the similar
OCs FSR\,31, FSR\,89 and FSR\,1744 (black triangles).}
\label{fig10}
\end{figure}

Finally, in Fig.~\ref{fig10} we show the spatial distribution in the Galactic plane of the
present FSR OCs, compared to that of the OCs in the WEBDA database. We consider the age ranges 
$\rm <0.4$\,Gyr, $\rm 0.4\,Gyr - 1$\,Gyr and $\rm >1$\,Gyr. FSR\,31, FSR\,89 and FSR\,1744
are also shown. Old OCs are found preferentially outside the Solar circle, and the inner Galaxy
contains few OCs so far detected. The interesting point here is whether inner Galaxy clusters
cannot be observed because of strong absorption and crowding, or have been systematically dissolved
by the different tidal effects combined (\citealt{OldOCs}, and references therein). In this context,
the more OCs are identified (with their astrophysical parameters derived) in the central parts, the
more constraints can be established to settle this issue.

\section{Summary and conclusions}
\label{Conclu}

The discovery of new star clusters in the Galaxy, and the derivation of their astrophysical parameters,
provide important information that, in turn, can be used in a variety of other studies related to the
star formation and evolution processes, dynamics of N-body systems, disruption time scales, the geometry
of the Galaxy, among others.

In this work we selected a sample of star cluster candidates projected nearly towards the dense stellar
field of the bulge ($|\Delta\ell|\la60^\circ$, $|\Delta\,b|\la20^\circ$), from the catalogue of \citet{FSRcat}.
They classified them as probable and possible star clusters, with quality flag '0' or '1'. The resulting 20
targets were analysed with 2MASS photometry by means of field-star decontaminated colour-magnitude diagrams,
colour-magnitude filters and stellar radial density profiles.

Of the 20 overdensities, 6 resulted with cluster-like CMDs and King-like RDPs (among these are the already
catalogued open clusters Harvard\,8=Cr\,268, and ESO\,275SC1). These are star clusters with ages in the range
0.6\,Gyr to $\sim5$\,Gyr, at distances from the Sun $\rm1.3\la\ds(kpc)\la2.8$, and Galactocentric distances
$\rm4.7\la\dgc(kpc)\la6.3$. Five others have CMDs and RDPs that suggest old star clusters, but they require
deeper photometry to establish their nature. Some of the uncertain cases might be globular clusters, considering
the high value of the field-star decontaminated CMD density parameter \ns\ and the similarity with the bulge CMD.
The remaining 9 overdensities are likely fluctuations of the associated dense stellar field.

Considering the above numbers, the fraction of overdensities that turned out to be star cluster ($f_{SC}$)
can be put in the range $30\%\la f_{SC}\la55\%$. The upper limit agrees with the $f_{SC}\approx50\%$
estimated by \citet{FSRcat}.

Systematic surveys such as that of \citet{FSRcat} are important to detect new star cluster
candidates throughout the Galaxy. Nevertheless, works like the present one, that rely upon field-star 
decontaminated CMDs and stellar radial profiles, are fundamental to probe the nature of such 
candidates, especially those projected against dense stellar fields.

\section*{acknowledgements}
We thank an anonymous referee for helpful suggestions. We acknowledge partial support 
from CNPq (Brazil). This research has made use of the WEBDA database, operated at the 
Institute for Astronomy of the University of Vienna.


\end{document}